%% file: Lucio.tex
\documentclass[submission,copyright,creativecommons]{eptcs}
\providecommand{\event}{PROLE 2016}
\usepackage[utf8]{inputenc}
\usepackage{amsmath}
\usepackage{amsfonts}
\usepackage{amssymb}
\usepackage{float}

\usepackage{listings}
\usepackage{dafny}
\usepackage[framemethod=tikz]{mdframed}
\usepackage{color}

\title{A Tutorial on Using Dafny to Construct Verified Software 
\footnote{
This work has been partially supported by the Spanish Project TIN2013-46181-C2-2-R and the Basque Project GIU15/30 and grant UFI11/45.}
}
\author{
Paqui Lucio  
\institute{The University of the Basque Country (UPV/EHU)}
\email{paqui.lucio@ehu.eus}
}
\def\titlerunning{Constructing Verified Programs with Dafny}
\def\authorrunning{Paqui Lucio}

\begin{document}

\maketitle

\begin{abstract}
This paper is a tutorial for newcomers to the field of automated verification tools, though we assume the reader to be relatively familiar with Hoare-style verification. In this paper, besides introducing the most basic features of the language and verifier Dafny, we place special emphasis on how to use Dafny as an assistant in the development of verified programs.
Our main aim is to encourage the software engineering community to make the move towards using formal verification tools.
\end{abstract}

\section{Introduction}
\label{intro}

The major goal of software engineers is to develop reliable software, that is to say error-free systems that accomplish the task for which they were designed.
In the relatively short history of software systems, several very well known software bugs with infamous (some even tragic) results, have been reported. 
The dramatic consequences of these errors have served as an incentive for  
further progress towards
more rigorous  (i.e. mathematical) software design, verification and
validation, in order to provide evidence of correctness and robustness of software.
On the one hand, many important firms --such as Microsoft, Ericsson and Rockwell Collins-- are increasing their investment in the development and use of technologies based on formal methods. On the other hand, the standards for certification of safety-critical software explicitly require the application of formal methods (\cite{GiP12}).
Software certification documents for aerospace, avionics, railway, medical devices, etc. promote the use of formal methods, emphasizing that critical software needs mathematical guarantees of its safety.
In particular, 
formal verification is mandated in the {\em Common Criteria for Information Technology Security
Evaluation} for the highest assurance level (cf. \cite{CoC12}).
The software engineering community has acquired substantial scientific knowledge, several formal methods, 
and a variety of technology for helping on the construction of trustworthy real systems.
Indeed, in the last decades, many successful examples of large-scale industrial software have been reported, from which we would like to mention the Sel4 microkernel \cite{KAE10,AJK12}, 
the Paris M\`etro Line 14 \cite{BBF99} and the Rotterdam Storm Surge Barrier \cite{MSE10}.
Moreover, the application of all the technology necessary to ensure reliability of safety-critical software may well be the major reason for
the absence, in the last 10 or 15 years, of serious damage caused by software failures.  
A traditional perception of formal verification as an expensive and difficult task to be applied to ``real-world” code could be provoking that only critical software is being verified.
Experiences with current tools
have demonstrated that this is not the case when the software is
implemented in a programming language
designed with software-reliability in mind.
With a minimum training in logic and some ability to understand and construct rigorous proofs,
the effort of learning and using a deductive verification tool is a justifiable investment.
Verification tools provide practical help to detect the presence or absence of bugs, thus yielding high-quality software.
There are many verification tools that can be used
in today's software development practices, such as Why3 \cite{FiP13} 
and KeY \cite{BHS07}. For a large, although incomplete, list of tools we refer to \cite{Bec14}.
The fact that automated verification is a research area
in continuous evolution enables the development of better and better practical tools.
Once we have the knowledge and technology to build up reliable software,  
why do not use them in the development of every software?
To do this, software engineers with expertise in formal methods is required and, to training them, software engineering curricula should provide foundations and training in the design and implementation of algorithms and data structures based on formal principles, abstract specifications and formal reasoning. 
In other words, automated program verification is already mature enough
to be made available in software industry. However, in the last decades and for different reasons, the software engineering curricula in some European universities (including Spanish ones) have experimented a drastic reduction contents and training related to mathematical and formal methods. In particular, formal verification of software has (totally or partially) disappeared from many curricula or has been made an elective subject. In addition, some fundamental subjects, such as formal logic, reasoning, semantics, etc. have undergone serious (even total) reductions. In our university there is  a traditional mandatory subject
that covers the topics of formal logic-based specification of programs, formal semantics based on the Hoare calculus and formal specification of data structures. In this subject, students must solve exercises (e.g. verify one-loop-programs, prove their termination, etc.) on paper.
Building on this basic course, our curriculum offers an elective subject on ``Formal Methods for Software Development" where we introduce tools for deductive software verification. In this elective subject, the main goal of the assignments is to produce automatically verified software, i.e. programs along with associated mathematical proofs of their properties, in particular their functional correctness.
The recent steady progress in the automation of program verification led us to think that an introductory primer on the use of these tools could be given in the middle of the mandatory course.
To learn how to use any particular tool, test-suite examples are very interesting, but sometimes the details of how the proofs were ``discovered" can be even more interesting and illustrating.
Finding a proof in an interactive verifier can be very
challenging. When a proof fails, the immediate feedback is helpful, but sometimes it is non-trivial to find out what help the verifier needs to make it through. Expertise in proof guidance is extremely important.
This paper is a tutorial that --besides introducing into the most basic features of Dafny-- places special emphasis on how to use Dafny as an assistant in the development of verified programs.
Indeed, it is intended to encourage newcomers to automated verification to start working with one of these tools, though we assume the reader to be relatively familiar with Hoare's logic.
We focus on the strategy for detecting and fixing problems while the annotated programs are constructed. For that, we deal with very simple programming examples.
This paper is not intended as a comprehensive review of Dafny, indeed we deal with a small subset of it.
In the official website of Dafny\footnote{\url{http://research.microsoft.com/en-us/projects/dafny/}} there is also an extensive online tutorial as well as a test-suite. In addition there are several papers covering a variety of applications and examples in Dafny, e.g. \cite{Lei10,Lei13b,Lei13a,LeL15}. 
All the annotated programs constructed in this tutorial can be found and tested in
\url{http://rise4fun.com/Dafny/r3pF}.

\section{Dafny and its Integrated Development Environment}

Dafny \cite{Lei10} 
is a program verification tool which includes a programming language
and specification constructs. The Dafny user creates and verifies both specifications and implementations.
The Dafny programming language is object-based, imperative, sequential, and
supports generic classes and dynamic allocation.
The specification constructs include standard pre- and postconditions, framing constructs,
and termination metrics. 
Dafny encourages using best-practice programming styles, in particular the {\em design by contract} approach.
Pre- and postconditions
state what properties have to be true at method entry and exit, respectively, and are used to construct the specifications or contracts of methods.
The method callers must stablish preconditions 
and can assume postconditions (after return).
The method implementers can assume preconditions and must stablish postconditions.
Dafny programs are statically verified for \emph{total correctness}, i.e., that every terminating execution satisfies its contract (\emph{partial correctness}) and that every execution does indeed terminate.
Dafny’s program verifier works by translating a given Dafny program into the \emph{intermediate verification language} Boogie \cite{BCD06} in such a way that the correctness of the Boogie program implies
the correctness of the Dafny program. Thus, the semantics of Dafny is defined in terms of Boogie. 
Boogie is a layer on which to build program verifiers for other languages. For example, the program verifiers VCC \cite{CDH09} for C, AutoProof for Eiffel \cite{AutoProof:TACAS2015}, and 
Spec\# \cite{SpecSharp:Retrospective:CACM}
are built on top of Boogie. 
The Boogie tool is used to generate first-order verification conditions that are passed to a logic reasoning engine. In particular, for Dafny, they are passed to the satisfiability modulo theories (SMT) solver Z3 \cite{MoN08}.
The Dafny integrated development environment (IDE) 
is an extension of Microsoft Visual Studio (VS). The IDE is designed to reduce the effort required by the user to make use of the proof system. For example, the IDE runs the program verifier in the background, thus providing design time feedback. 
It makes Dafny to be an auto-active verifier \cite{Lei13b}: users will interact with the proofs, but there is automation. Specification and programming constructs can be used to create the proof.
Dafny allows declarative calculations and inductive
as well as co-inductive proofs.
Also, verification error messages can have a lot of associated information 
and the user can get information about the possible values of variables for a reported error using the Boogie Verification Debugger (BVD) \cite{LLM11} that is deeply integrated into the Dafny IDE. 
The interested reader is referred to \cite{LeW14} for further information on the Dafny IDE.

\section{A First Example of Annotated Program}
\label{basic}

In this section, we will start designing an annotated program to compute the factorial of a natural number. We intentionally use a {\em non-typical} invariant to obtain a verified algorithm that is different from the usual one, with 
the main aim of introducing several basic concepts and units in Dafny.

The basic unit of a Dafny program is the \D{method}. A method is a piece of executable code with a head where multiple named parameters and multiple named results are declared.
As a first example we design an iterative method that calculates the factorial of a natural number from a very peculiar invariant aiming to provide interesting insights.
Dafny also offers user-defined \D{function}s.
By default in Dafny, functions can be used only in specifications, hence they do not generate code. To override this default, so that the compiler will generate code for a function, the function is declared with \D{function method}. 
A \D{predicate} is a boolean function, and a \D{predicate method} is a predicate for which code is generated. Dafny has built-in specification constructs for assertions, such as \D{requires} for preconditions, \D{ensures} for postconditions, \D{invariant} for loop invariants, \D{assert} for inline assertions. 
Multiple \D{requires} have the same meaning as their conjunction into a single \D{requires}. \footnote{The same will apply to multiple \D{ensures}, \D{invariant}, and \D{assert}.}
The starting point is
\begin{lstlisting}[numbers=left, stepnumber=3, firstnumber=1]
function factorial (n:int): int
  requires n rel>= 0;
{
if n == 0 then 1 else n * factorial(n-1)
}

method computeFactorial(n:int) returns (f:int)
  requires n rel>= 0
  ensures f == factorial(n)
// annotated code will be designed from the invariant 
// 0 rel<= i rel<= n-1 && f * factorial(n-i-1) == factorial(n)
\end{lstlisting}
where we have used a function \D{factorial} for specification purposes: it is used in the postcondition of the method and also in the intended invariant, which is a rather unusual one.
The precondition of the function  \D{factorial} serves to ensure the well-foundedness of the induction. The interested reader can check the error that Dafny reports if you erase/comment the above line 2.
However, with the precondition \D{n rel>= 0}, Dafny guesses that \D{n} is a termination metric, then the hover text \D{decreases n} appears by puting the cursor over the word \D{factorial}. That is, \D{n} is the expression whose strict decrease warrants the recursion termination.
We will discuss that topic in Section \ref{termination}.

If we would like to generate code for the function \D{factorial} we also could declare it as follows: 
\D{function method factorial (n:int): int}
with the same requires clause and the same body.
In that case, the result of the function method is not named, whereas in the method \D{computeFactorial} the result is named as \D{f}. They both also use two different algorithms to compute the same result:

\hspace{-8mm}
\includegraphics[scale=0.65]{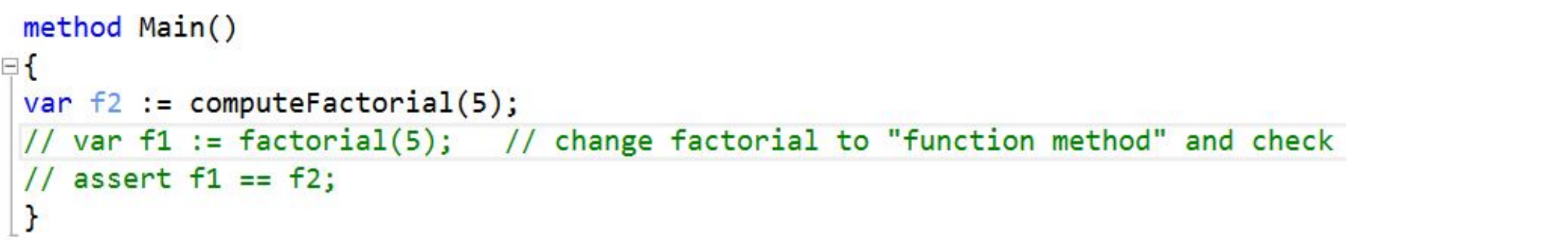} \\
According to the invariant \D{0 rel<= i rel<= n-1 &&  f * factorial(n-i-1) == factorial(n);} the auxiliary variable \D{i} and the result \D{f} should be initialized by \D{0} and \D{n}, respectively. It is also easy to check that (in the iteration) after an assignment \D{i := i + 1} there should be some assignment to \D{f} satisfying 

\D{0 rel<= i rel<= n-1 &&  f * factorial(n-i) == factorial(n);} 

\D{f := ?}

\D{0 rel<= i rel<= n-1 &&  f * factorial(n-i-1) == factorial(n);} \\
In this way we deduce that \D{f} should be assigned to be \D{f * (n-i)}.
The resulting program is not yet correct.
Indeed an error is displayed by a red dot, a red curly underline and the associated hover text (which appears when the cursor is placed over the red wavy line):\\
\includegraphics[scale=0.35]{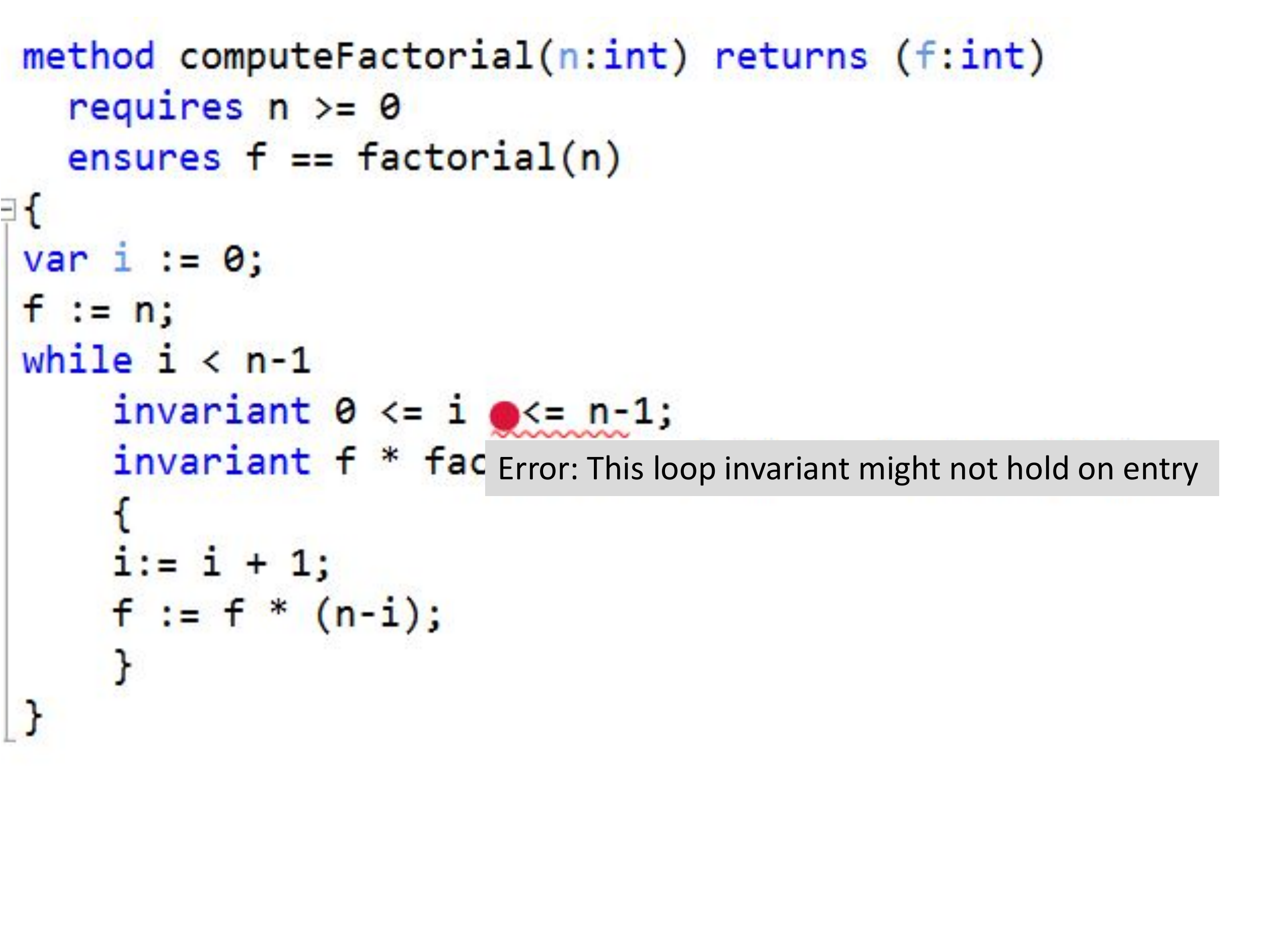} \\[-1.3cm]
Indeed, the case \D{n==0} is in conflict with the assertion.
A suitable fixing is
\begin{lstlisting}[numbers=left, stepnumber=3, firstnumber=1]
method computeFactorial(n:int) returns (f:int)
  requires n rel>= 0
  ensures f == factorial(n)
{
var i := 0;
if n == 0 { f:= 1;}
    else {
         f := n;
         while i rel< n-1
            invariant 0 rel<= i rel<= n-1;
            invariant f * factorial(n-i-1) == factorial(n);
            {
            i:= i + 1;
            f := f * (n-i);
            }
         }
}
\end{lstlisting}
Now, we are going to design the same program in a different way to illustrate how to use a specified (but not still implemented) method. This enables a kind of modular design by means of the fact that when a method \D{M} calls a method \D{M'}, only the contract of \D{M'} (but not its body/code) is required in the correctness proof of the method \D{M}. For example, we can abstract the body of the iteration using a call to a lemma \D{oneStep} with contract

\begin{lstlisting}
method oneStep(i:int,f:int,n:int) returns (i':int,f':int)
   requires 0 rel<= i rel< n-1 && f * factorial(n-i-1) == factorial(n)
   ensures 0 rel<= i' rel<= n-1 && f'* factorial(n-i'-1) == factorial(n)
\end{lstlisting}
in this way the assignment \D{i,f := oneStep(i,f,n);} might work as the previous body of two assignments. However, we get the following error: \footnote{We have moved up the hover text for clarity.}

\vspace{-6mm}
\hspace{-9mm}
\includegraphics[scale=0.4]{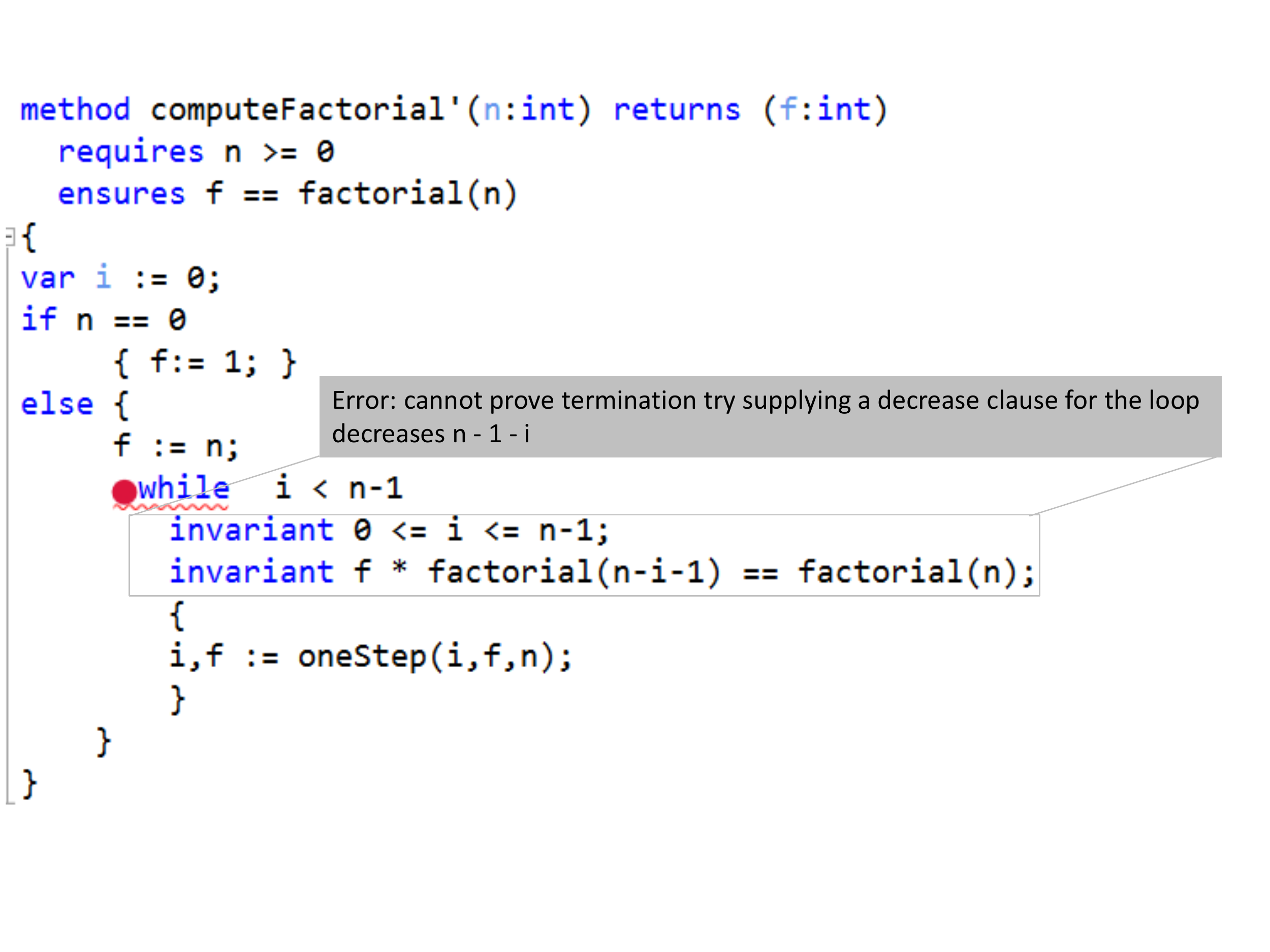} \\[-1.2cm]
Though Dafny guesses that the expression that decreases at each step could be \D{n-1-i}, it is not able to prove that fact. 
Indeed, the contract of the method \D{oneStep} --the only available information-- does not state anything that is related to the modification of the variables affecting the termination and that could be used to prove the corresponding verification conditions.
Therefore, an additional \D{ensures n-i' rel< n-i}, or simply
\D{ensures i' rel> i}, fixes the problem.
To complete this {\em modular version} of a method for computing the factorial we should write a body for the method \D{oneStep} with the two assignments bellow:
\begin{lstlisting}[numbers=left, stepnumber=3, firstnumber=1]
method oneStep(i:int,f:int,n:int) returns (i':int,f':int)
	requires 0 rel<= i rel< n-1 && f * factorial(n-i-1) == factorial(n)
	ensures 0 rel<= i' rel<= n-1 && f'* factorial(n-i'-1) == factorial(n)
	ensures i' rel> i
{
i':= i + 1;
f' := f * (n-i');
}
\end{lstlisting}
Now, the error is repaired and the hover text over the word \D{while} of the method \D{computeFactorial}  reads \D{decreases n-1-i}, but the interested user can write (next to the invariant) the annotation \D{decreases n-i} or any other correct expression which is strictly decreasing. Doing so, Dafny checks the verification conditions for the user-expression, instead of for the guessed one.
We will go back to this example in Section \ref{termination} where we discuss more on termination.

\section{Calling lemmas in program proofs}

In this section, we illustrate how we guess and use lemmas that are required to verify a program. 
First, we design a verified method that computes the function $2^{3k}-3^k$ which is divisible by 5, 
hence it computes $5*f(k)$ where $f(k) = (2^{3k}-3^k) / 5$. We start specifying the method as follows

\begin{lstlisting}[numbers=left, stepnumber=3, firstnumber=1]
method compute5f (k:int) returns (r:int)
  requires k rel>= 1
  ensures r == 5*f(k) 

function f(k:int):int
  requires k rel>= 1;
{  (exp(2,3*k) - exp(3,k)) / 5  }

function exp(x:int,e:int):int
  requires e rel>= 0 
{ if e==0 then 1 else x * exp(x,e-1) }
\end{lstlisting}
Then, we design an iteration that calculates both exponentials $2^{3n}$ and $3^n$ by iterating multiplication by $8$ and by $3$, respectively, in two new variables \D{t1} and \D{t2}. That is, an iteration that preserves the invariant
\D{0 rel<= i rel<= k && t1 == exp(2,3*i) && t2 == exp(3,i)}
where \D{i} is also a new variable.

\begin{lstlisting}[numbers=left, stepnumber=3, firstnumber=1]
method compute5f (k:int) returns (r:int)
  requires k rel>= 1
  ensures r == 5 * f(k) 
{
var i, t1, t2:= 0, 1, 1;
while i rel< k
	invariant 0 rel<= i rel<= k;
	invariant t1 == exp(2,3*i);
	invariant t2 == exp(3,i);
	{
	i, t1, t2 := i+1, 8*t1, 3*t2;
	}
r := t1-t2;
}
\end{lstlisting}
%\vspace{-5mm}
Dafny complains that the (unique) postcondition does not hold and also that the second conjunct of the provided invariant might not be maintained by the loop:

\hspace{-8mm}
\includegraphics[scale=0.6]{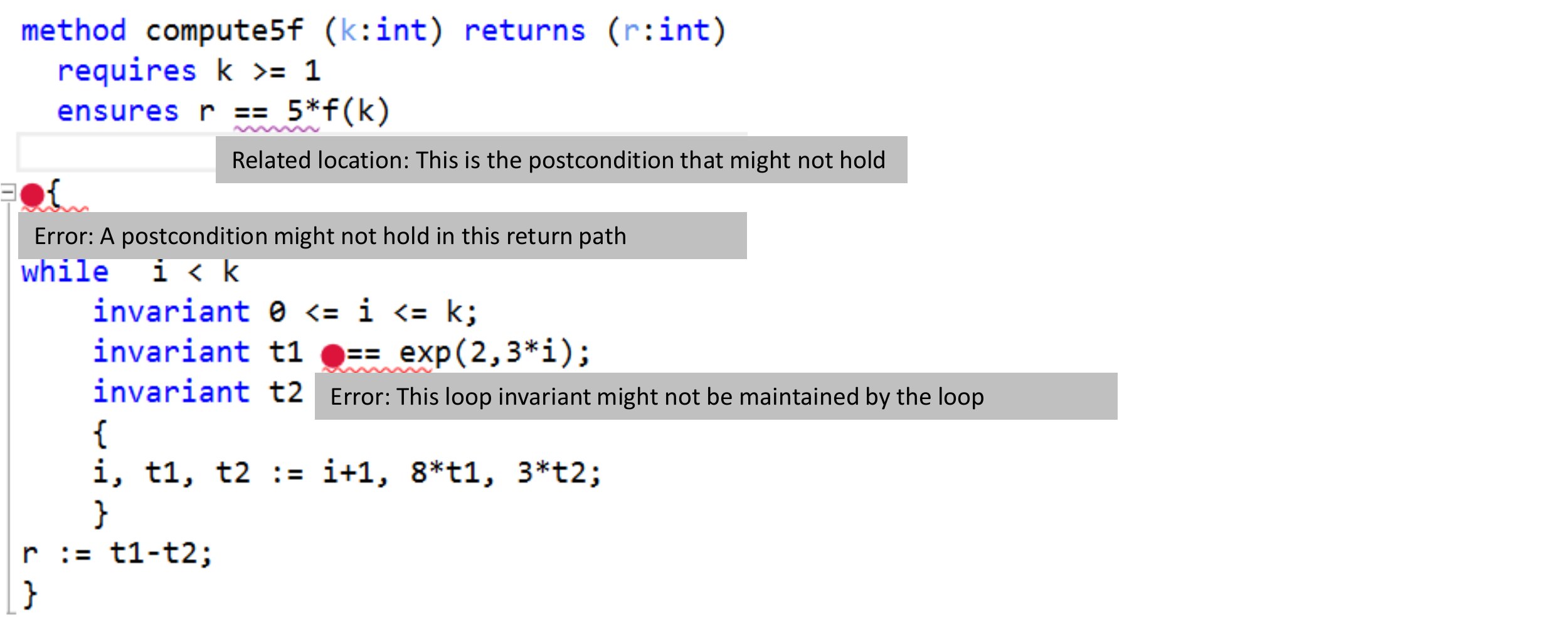} \\ 
Dafny cannot complete the proof (that is, the SMT-solver Z3 cannot prove all verification conditions) by itself. Then, 
the user should provide assertions as hints. A {\em hint} is an assertion that the verifier is required to prove. Once the assertion is proved, it turns into a usable property for completing the correctness proof.
Indeed, ``\D{assert} $\varphi$" tells Dafny to check that $\varphi$ 
holds (whenever control reaches that part of the code) and to use the condition $\varphi$ (as a lemma) to prove the  verification conditions beyond this program point. In order to help the user in the process of constructing proofs, in particular in the process of {\em guessing hints}, Dafny offers two features: the construct \D{assume} and the use of a declared (but yet not proved) \D{lemma}.
Therefore, along the construction of a proof, we can introduce an assumed condition $\varphi$ to check whether $\varphi$ is the condition that Dafny needs to complete the proof. In other words, Dafny tries to complete the proof, assuming that $\varphi$ is true, without having tried to prove $\varphi$.
%In general, invariants are needed to prove that postconditions hold. For that, the user should provide a loop invariant that should be held on entry, should be maintained by the loop, and should be strong enough to ensure the postcondition.\\
In order to check whether the proposed invariant is suitable for ensuring the postcondition, we write \D{assume t1 == exp(2,3*i);} in line 12 and the assert of line 15 is not violated.
Since the postcondition is not yet held, we add the assumption in line 16:

\begin{lstlisting}[numbers=left, stepnumber=3, firstnumber=1]
method compute5f (k:int) returns (r:int)
  requires k rel>= 1
  ensures r == 5*f(k) 
{
var i, t1, t2:= 0, 1, 1;
while i rel< k
	invariant 0 rel<= i rel<= k;
	invariant t1 == exp(2,3*i);
	invariant t2 == exp(3,i);
	{
	i, t1, t2 := i+1, 8*t1, 3*t2;
	assume t1 == exp(2,3*i);
	}
r := t1-t2;
assert r ==  exp(2,3*k) - exp(3,k);
assume r == 5 * f(k);
}
\end{lstlisting}
When Dafny succeeds with an assumption ``\D{assume} $\varphi$", then it should be changed to ``\D{assert} $\varphi$" to force Dafny to prove $\varphi$.
If $\varphi$ is proved, then the inline assertion is the required hint.
In our example, we should help Dafny to prove the invariant-preservation verification condition related to the second conjunct of the invariant.
Hence, we convert the \D{assume} in the above line 12 into an \D{assert} (see line 14 below)
and, as expected, the assertion is violated. Therefore we use the weakest precondition technique to write 
\D{assume 8*t1 == exp(2,3*(i+1));} (see line 11 below).
Then, the assertion in line 13 (below) is verified. 
\begin{lstlisting}[numbers=left, stepnumber=3, firstnumber=1]
method compute5f (k:int) returns (r:int)
  requires k rel>= 1
  ensures r == 5*f(k) 
{
var i, t1, t2:= 0, 1, 1;
while i rel< k
	invariant 0 rel<= i rel<= k;
	invariant t1 == exp(2,3*i);
	invariant t2 == exp(3,i);
	{
	assume 8*t1 == exp(2,3*(i+1));
	i, t1, t2 := i+1, 8*t1, 3*t2;
	assert t1 == exp(2,3*i);
	}
r := t1-t2;
assume (exp(2,3*k) - exp(3,k)) % 5 == 0;
assert r == 5 * f(k);
}
}
\end{lstlisting}
For helping with the postcondition, when we swap \D{assume r == 5 * f(k)} to an assertion, it is violated. Since the assert \D{assert r ==  exp(2,3*k) - exp(3,k)}, is satisfied (not violated) we guess that Dafny ``does not know" that \D{exp(2,3*k) - exp(3,k)} is divisible by 5. Therefore, we write the assumption in line 17 below.
But now, if we change 
both \D{assume} clauses (in lines 11 and 16) to \D{assert}s, both are  violated.
In line 11, we try to help Dafny by unfolding the assertion \D{ 8*t1 == == exp(2,3*(i+1))} into
\begin{center}
\D{assert 8*t1 == 8*exp(2,3*i) == exp(2,3*i+3) == exp(2,3*(i+1))}
\end{center}	
Then, the prover fails checking the equation \D{8*exp(2,3*i) == exp(2,3*i+3)}. This is shown by a red point on the equality symbol. Similarly, in line 16, a red point is in its unique equality symbol.
Then, we try to help Dafny to prove both properties: \D{exp(2,3*i)*8 == exp(2,3*i+3)} and \D{(exp(2,3*k) - exp(3,k)) % 5 == 0}.
If Dafny could prove them, then Dafny would respectively use them to prove the postcondition and the invariant preservation, so that these assertions would be the required hints. Unfortunately, the assertions cannot be used as hints, but must be proved as separated lemmas.
Hence, we write both lemmas and call them as follows:
%\begin{figure}[H]
\begin{lstlisting}[numbers=left, stepnumber=3, firstnumber=1]
method compute5f (k:int) returns (r:int)
  requires k rel>= 1
  ensures r == 5*f(k) 
{
var i, t1, t2:= 0, 1, 1;
while i rel< k
	invariant 0 rel<= i rel<= k;
	invariant t1 == exp(2,3*i);
	invariant t2 == exp(3,i);
	{
	expPlus3_Lemma(2,3*i);
	// assert t1*8 == exp(2,3*i)*8 == exp(2,3*i+3) == exp(2,3*(i+1));
	i, t1, t2 := i+1, 8*t1, 3*t2;
	}
r := t1-t2;
DivBy5_Lemma(k);
// assert (exp(2,3*k) - exp(3,k)) % 5 == 0;
}

lemma expPlus3_Lemma (x:int,e:int)
	requires e rel>= 0;
	ensures x * x * x * exp(x,e) == exp(x,e+3);
// to be proved

lemma DivBy5_Lemma (k:int)
	requires k rel>= 1
	ensures (exp(2,3*k) - exp(3,k)) % 5 == 0
// to be proved
\end{lstlisting}
%\vspace{-5mm}
%\caption{Two lemmas helping in a program verification}
%\label{fig-two-lemmas}
%\end{figure}
We use each lemma call to validate the respective \D{assert}. That is, if a concrete lemma call  works (in the sense that $\varphi$ is satisfied) then we could comment (or drop) the assert.
The interested reader could uncomment the above lines 12 and 19 to check out that the lemma call ensures the non-violation of the assertion. It is useful to keep commented assertions as explanations or program documentation.

The \D{lemma} declarations are like methods but do not need to be called at runtime, hence no code is generated for them, i.e.  \D{lemma} is equivalent to \D{ghost method}.\footnote{There are more ghost entities in Dafny (see Section \ref{termination})}
Lemmas have only effect on the verification of the program, they are only necessary to help along the proof of the program. Lemmas, like methods, can have parameters, hence their re-usability by instantiation (of the parameters) is an advantage of lemmas with respect to inline asserts. In particular, the induction hypothesis of inductive lemmas is invoked as a lemma call (over smaller parameters).
In Dafny, a lemma is a ghost method whose contract represents the property it warrants. Let us consider any lemma
\begin{lstlisting}[numbers=left, stepnumber=5, firstnumber=1]
lemma Ex_Lemma (x1:T1,...,xn:Tn)
  requires \varphi
  ensures \psi
{ body }
\end{lstlisting}
where \D{x1,...,xn} is the tuple of formal parameters and \D{T1,...,Tn} the tuple of respective types.
The contract of lemma \D{Ex_Lemma} means \D{forall x1,x2,...,xn :: \varphi ==> \psi}, and its body is a proof of such property.
A lemma call like  \D{ Ex_Lemma(a) }
where \D{a} is the tuple of current parameters corresponds  to 
\begin{lstlisting}
assert \varphi[a/x];
assume \psi[a/x];
\end{lstlisting}
That is, if the precondition --for the current parameters \D{a}-- can be proved, we can assume that the postcondition holds --for the current parameters \D{a}. The assume clause is discharged whenever the lemma is proved.

A program-proof is not complete until all verification conditions have been discharged, i.e., all assume
statements have been removed (or replaced by asserts), and all the lemmas have been proved. Hence, in our example, we should prove the lemmas, i.e. we should write their body.

\section{Proving lemmas}

In this section we focus on proving the two lemmas left unproved in the previous section. 
Sometimes Dafny is able to prove a lemma (even an inductive lemma) by itself. This means that the empty body (i.e. $\{  \}$) serves as proof. Otherwise, lemma proofs are annotated code as the body of an executable (non-ghost) method. In particular, we can use properties as hints in lemma bodies.
For example, for the following lemma (from the previous section) the empty body does not work:
\begin{lstlisting}
lemma expPlus3_Lemma (x:int,e:int)
  requires e rel>= 0;
  ensures x * x * x * exp(x,e) == exp(x,e+3);
\end{lstlisting}
It is easy to see that it can be proved by means of three successive applications (or unfoldings) of the recursive definition of the function \D{exp}. Hence, we try to give just one hint as lemma body:
\begin{lstlisting}
{
assert  x * x * x * exp(x,e) == x * x * exp(x,e+1) == x * exp(x,e+2) == exp(x,e+3);
}
\end{lstlisting}
and we succeed. Indeed also 
\begin{lstlisting}
{
assert  x * exp(x,e) == exp(x,e+1);
}
\end{lstlisting}
serves as hint, since Z3 automatically performs the other two unfoldings. 

Dafny provides a special notation that is easy to read and understand: {\em calculations} \cite{Lei14}.
 A calculation in Dafny is an statement that proves a property.
This notation was extracted from the  {\em calculational method} \cite{Bac95}, whereby a theorem is established
by a chain of formulas, each transformed in some way into the next.
The relationship between successive formulas (for example, equality, implication, double implication, etc.) is notated, or it can be omitted if it is the default relationship (equality). In addition, the hints (usually asserts or lemma calls) that justify a step can also be notated
(in curly brackets after the relationship).
Calculations are written inside the environment \D|calc{ }|.
The grammar for calculations is:

\begin{lstlisting}
CalcStatement := calc { 
                      CalcBody 
                      }
CalcBody := Line 
           (Op Hint 
            Line)*        
Line := Expression ;
Op :=  rel<= | rel< | rel>= | rel> | ==> | c<== | c<==> | != | ==
Hint := { (BlockStatement | CalcStatement)* }
\end{lstlisting}
where a \D{BlockStatement} is one or more assert clauses (or provisionally assume clauses), lemma calls and forall-statements. 

Like in other proof assistants (e.g. Isabelle/HOL and Coq) and verifiers (e.g. Why3 and KeY), Dafny 
allows proofs to be written in different styles and with different levels of description of the outcome of every logical transformation.
Therefore, proof readability and easy checking by humans is part of the work of the Dafny user. To illustrate that issue we will give two different proofs for the property \D{DivBy5_Lemma}. The first proof shows a pedantic number of details:
\begin{lstlisting}[numbers=left, stepnumber=3, firstnumber=1]
lemma DivBy5_Lemma (k:int)
  requires k rel>= 1
  ensures (exp(2,3*k) - exp(3,k)) % 5 == 0
{
if k==1 {
       // the base case is automatically proved
       }
  else {
       calc {
            (exp(2,3*k) - exp(3,k)) % 5;
            ==
            (exp(2,3*(k-1)+3) - exp(3,(k-1)+1)) % 5;
            == {
               expPlus3_Lemma(2,3*(k-1));                   // lemma call
              }
            (exp(2,3*(k-1))*8 - exp(3,(k-1))*3) % 5;
            ==
            (3*( exp(2,3*(k-1)) - exp(3,k-1) ) + 5* exp(2,3*(k-1))) % 5;
            == {
               DivBy5_Lemma(k-1);                     // lemma call for IH
               // assert (exp(2,3*(k-1)) - exp(3,k-1)) % 5 == 0;      // IH
               } 
            0;
            }
         }
}
\end{lstlisting}
The above proof is divided into two cases, as indicated by an if statement.
The base case, for
k=1, is automatically proved in lines 5-7. The other case uses a calculational proof that proves that the expression \D{exp(2,3*k) - exp(3,k)} is a multiple of \D{5}. In the calculation we use two lemma calls as hints in lines 14 and 21. The latter calls this lemma itself recursively. This call is treated in accordance
with programming rules: the precondition of the callee is checked, termination is checked, and then the postcondition can be assumed.
In effect, this sets up a proof by induction, where the recursive call to the lemma obtains
the inductive hypothesis. 
Next, an alternative proof with only
the essential hints:

\begin{lstlisting}[numbers=left, stepnumber=3, firstnumber=1]
lemma DivBy5_Lemma_Simplified(k:int)
  requires k rel>= 1
  ensures (exp(2,3*k) - exp(3,k)) % 5 == 0
{
if k>1 {
       expPlus3_Lemma(2,3*(k-1));       // lemma call
       DivBy5_Lemma_Simplified(k-1);    // lemma call for IH	
      }
}
\end{lstlisting}
A user can decide to keep as many details as are
deemed helpful for human understanding and to elide those that seem more like clutter.
The second proof may take more head-scratching for a human to understand.
Provided enough hints are supplied for Dafny to complete the proof, the trade-off
between clarity and clutter is up to the user and depends on how many details the user
wants to show for human readers.

\section{A verified bubble sort method}

In this section we explain a verified implementation of the well-known bubble sort algorithm. The aim of this example is twofold. It gives us the opportunity to introduce predicates, arrays, mutable states and framing in Dafny, whereas we detail a more interesting example of modular design based on the design-by-contract paradigm. 

To start, we should define a predicate \D{sorted} whose parameter is an array of integers and a method \D{bubbleSort} as follows:
\begin{lstlisting}[numbers=left, stepnumber=3, firstnumber=1]
predicate sorted (a:array<int>)
	requires a != null
	reads a

method bubbleSort (a: array<int>)
	requires a != null
	modifies a
	ensures sorted(a)
\end{lstlisting}
Arrays are a built-in part of the language Dafny, with their
own type \D{array<T>}, where \D{T} is another type.
Arrays are objects, hence mutable, that are dynamically allocated, hence they can be null. We write \D{requires a != null} to state that both the predicate \D{sorted} and the method \D{bubbleSort} cannot be applied to a null object. For dealing with a mutable state, every method specification should say which objects are allowed to
be changed, every function (in particular predicate) specification should detail the set of mutable objects which it depends on.
This is called {\em framing}: the former is specified by a \D{modifies} clause, like the one in the above method \D{bubbleSort}
(see line 7) and the latter by a \D{reads} clause, like the one in the predicate \D{sorted} (see line 3).
So, \D{reads} and \D{modifies} clauses provide a set of objects.
Methods are allowed to read whatever they want, so their
\D{reads} clauses do not need to be specified.
Functions/Predicates are not allowed to modify objects, so
\D{modifies} clauses cannot be specified.
The \D{modifies} clause says that the method is allowed to
modify the state of any of those objects.
The \D{reads} clause says that the function is allowed to depend
on the state of any of those objects.

A complete specification of any sorting method should say that the final object state is a permutation of the initial object state, but we intentionally delay the permutation part of the specification until the code with the sorting property has been verified. Of course, our code will only swap elements in the array.

By now, we can define the predicate \D{sorted}  by
\D{forall i,j :: 0 rel<= i rel< j rel< a.Length ==> a[i] rel<= a[j]}.

Then, we have to write annotated code for the method
\D{bubbleSort} which we would like it to work as shown in the following example:

\hspace{-8mm}
$
\begin{array}{cccccc}
\mbox{Step}\; i=1 & 7, \underline{2}, 6, 3, 4 & \Longrightarrow & \underline{2},\underline{7}, 6, 3, 4 & &\\
\mbox{Step}\; i=2 & 2, 7, \underline{6}, 3, 4 & \Longrightarrow & 2, \underline{6}, \underline{7}, 3, 4 & &\\
\mbox{Step}\; i=3 & 2, 6, 7, \underline{3}, 4 & \Longrightarrow & 2, 6, \underline{3}, \underline{7}, 4
  & \Longrightarrow & 2, \underline{3}, 6, \underline{7}, 4\\
\mbox{Step}\; i=4 & 2, 3, 6, 7, \underline{4}
 & \Longrightarrow &  2, 3, 6, \underline{4}, \underline{7}
  & \Longrightarrow &  2, 3, \underline{4}, 6, \underline{7}\\
\end{array}
$\\
Hence, we get the following first version of annotated code:\footnote{where we have moved up the hover text for visibility.}

\vspace{-4mm}
\hspace{-6mm}
\includegraphics[scale=0.41]{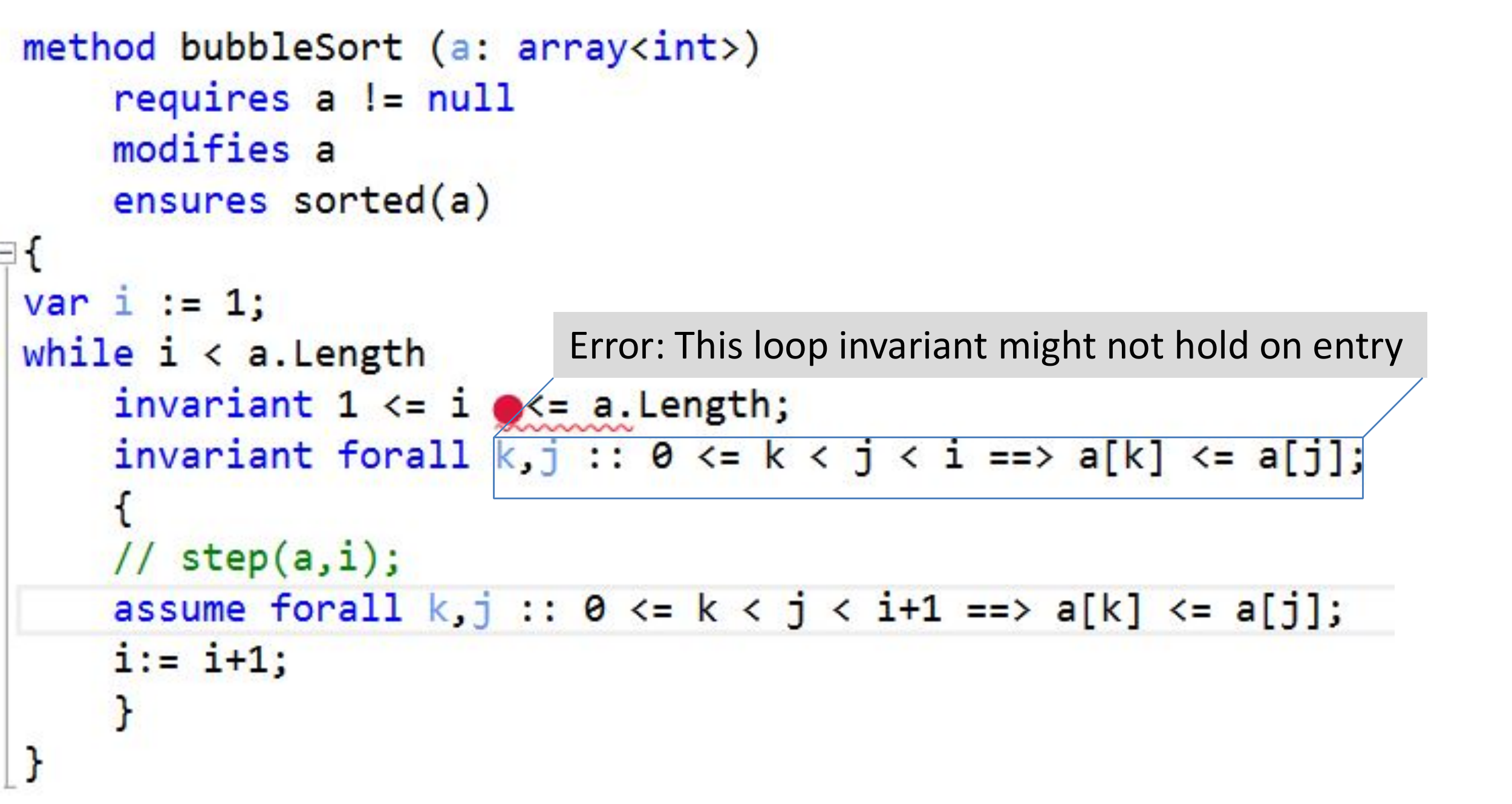} \\
It is easy to realize that \D{ i rel<= a.Length } cannot be warranted at the loop entry since \D{i := 1} is the loop initialization, unless \D{i rel> 1} holds in entry. We can add a precondition \D{a.Length rel> 1}, the interested reader could check that it works. Since any singleton array is sorted, a better thing to do is to adapt code to do nothing with singletons. At the same time, we realize that the sorted property is used in the precondition, in the invariant, in the assume clause, and will be used in the not yet defined method \D{bubbleStep}.
So it would be useful to parameterize the predicate \D{sorted}. Consequently, we get 
the new version:

\begin{lstlisting}[numbers=left, stepnumber=3, firstnumber=1]
predicate sorted (a:array<int>)
  requires a != null
  reads a
{
sortedBetween(a,0,a.Length)
}

predicate sortedBetween (a:array<int>, lo:int, hi:int)
  requires a != null && 0 rel<= lo rel<= hi rel<= a.Length
  reads a
{
forall i,j :: lo rel<= i rel< j rel< hi ==> a[i] rel<= a[j]
}

method bubbleSort (a: array<int>)
  requires a != null
  modifies a
  ensures sorted(a)
{
if a.Length rel> 1
	{
	var i := 1;
	while i rel< a.Length 
		invariant 1 rel<= i rel<= a.Length;
		invariant sortedBetween(a,0,i);
		{
		// bubbleStep(a,i);
		assume sortedBetween(a,0,i+1);
		i:= i+1;
		}
	}
}
\end{lstlisting}
Then, we can comment the assume clause in line 29 and uncomment the lemma call \D{bubbleStep(a,i)} in line 28 whenever we add the following specification of the method \D{bubbleStep}:
\begin{lstlisting}[numbers=left, stepnumber=3, firstnumber=35]
method bubbleStep (a: array<int>, i:int)
  requires a != null && 0 rel<= i rel< a.Length && sortedBetween(a,0,i) 
  modifies a
  ensures sortedBetween(a,0,i+1)
\end{lstlisting}
Next, annotated code for the latter method is required. 
Consider the last step for $i=4$:

\vspace{-4mm}

$$
\begin{array}{ccccc}
2, 3, 6, 7, \underline{4}
 & \Longrightarrow &  2, 3, 6, \underline{4}, \underline{7}
  & \Longrightarrow &  2, 3, \underline{4}, 6, \underline{7}\\
\end{array}
$$
%\vspace{-4mm}
It is easy to see that a new variable \D{j} should be initialized to be \D{i} and should be decreased while \D{a[j-1] rel> a[j]} whereas these two elements are swapped. It is also easy to see that such process preserves the invariant
\D{sortedBetween(a,0,j) && sortedBetween(a,j,i+1)}, but all that is not still correct:\\

\vspace{-2mm}
\hspace{-8mm}
\includegraphics[scale=0.52]{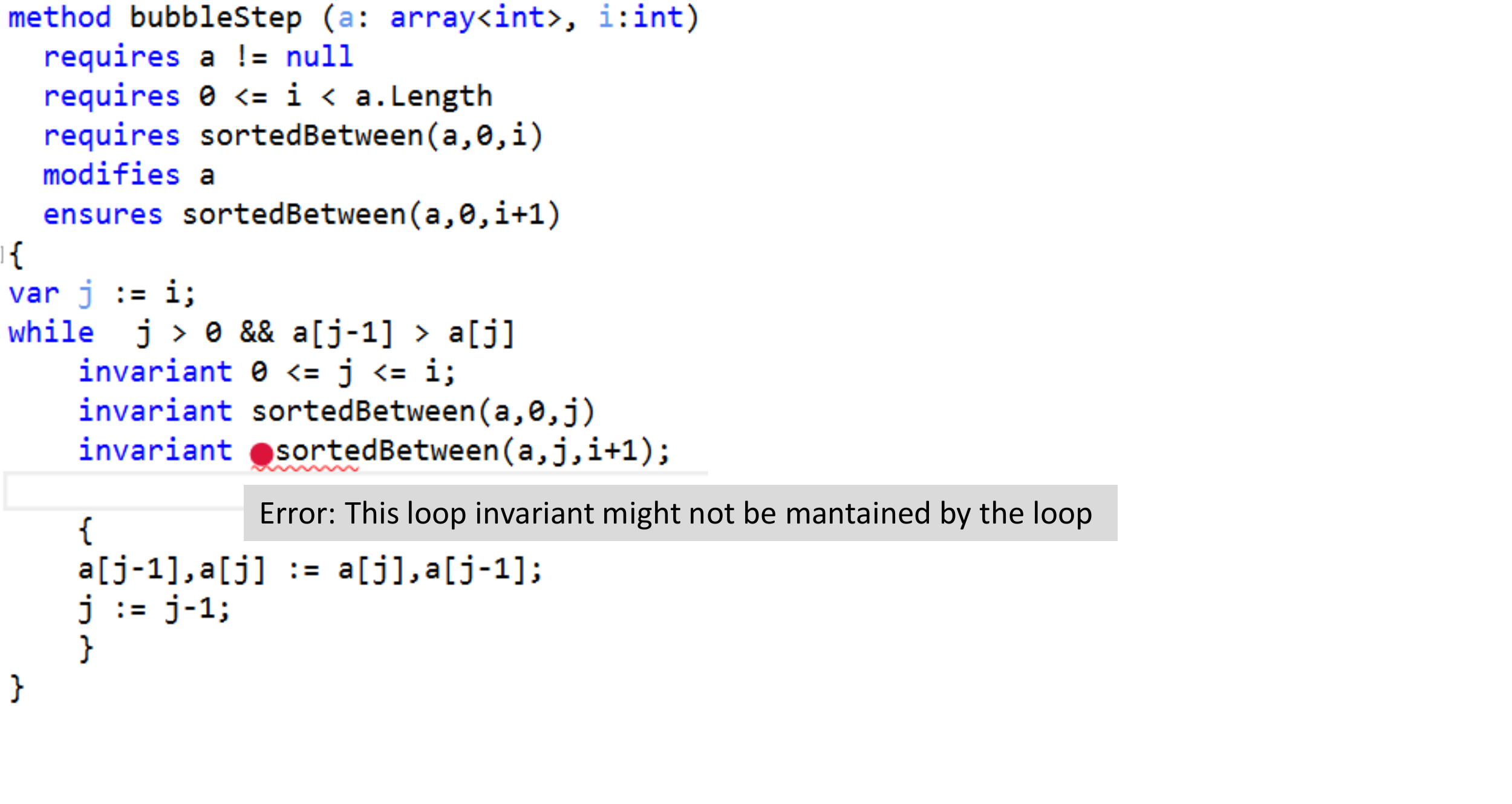}\\[-7mm]
Dafny can help us to analyze why this invariant might not be maintained by the loop. We use the previously explained assume/assert mechanism to guess the assumption in the line 15 below: 
\begin{lstlisting}[numbers=left, stepnumber=3, firstnumber=1]
method bubbleStep (a: array<int>, i:int)
  requires a != null
  requires 0 rel<= i rel< a.Length
  requires sortedBetween(a,0,i) 
  modifies a
  ensures sortedBetween(a,0,i+1)
{
var j := i;
while j rel> 0 && a[j-1]  rel> a[j] 
	invariant 0 rel<= j rel<= i;
	invariant sortedBetween(a,0,j) 
	invariant sortedBetween(a,j,i+1);
	{
	//assert a[j-1] rel> a[j] && sortedBetween(a,j,i+1);
	assume j+1 rel<= i ==> a[j-1] rel<= a[j+1];
	a[j-1],a[j] := a[j],a[j-1];
	assert sortedBetween(a,j-1,i+1);
	j := j-1;
	assert sortedBetween(a,j,i+1);
	}
} 
\end{lstlisting}
We first assume the property in line 19 and, of course, everything works, but as an assertion (line 19) it is violated. Then, we fix this problem by assuming the property in line 17. Again, as an assertion line 17 is also violated. Now, we calculate that
\D{a[j] rel<= a[j-1] rel<= a[j+1] rel<=} \D{... rel<= a[i] rel<= a[i+1]} should be true  in line 15.
Since the commented assertion in line 14 is true, we guess that we must assume that
\D{a[j-1] rel<= a[j+1]} with the guard \D{j+1 rel<= i} for preventing 
``Error: index out of range". 
Again, changing that property to an assert, we get an ``Error: assertion violation"
Hence, we learn that the assumed property should be  part of the invariant. Adding \D{invariant j+1 rel<= i ==> a[j-1] rel<= a[j+1]}
produces an out-of-range error on index \D{j-1} because \D{j} is \D{0} at the end of the loop (but not inside the loop). Consequently, 
the invariant to be added is:
 \D{1 rel< j+1 rel<= i ==> a[j-1] rel<= a[j+1]}.

\begin{figure}[H]
\begin{lstlisting}[numbers=left, stepnumber=3, firstnumber=1]
predicate permutation (a:seq<int>,b:seq<int>)
{  multiset(a) == multiset(b)  }

predicate sortedBetween (a:array<int>, lo:int, hi:int)
  requires a != null && 0 rel<= lo rel<= hi rel<= a.Length
  reads a 
{  forall i,j :: lo rel<= i rel< j rel< hi ==> a[i] rel<= a[j] }

predicate sorted (a:array<int>)
  requires a != null
  reads a
{ sortedBetween(a,0,a.Length) }

method bubbleSort (a: array<int>)
  requires a != null
  modifies a
  ensures sorted(a) 
  ensures permutation(a[..],old(a[..]))
{
if a.Length  rel> 1
	{
	var i := 1;
	while i rel< a.Length 
		invariant 1 rel<= i rel<= a.Length;
		invariant sortedBetween(a,0,i);
		invariant permutation(a[..],old(a[..]));
		{
		bubbleStep(a,i); 
		i:= i+1;
		}
	}
}

method bubbleStep (a: array<int>, i:int)
  requires a != null && 0 rel<= i rel< a.Length && sortedBetween(a,0,i) 
  modifies a
  ensures sortedBetween(a,0,i+1) 
  ensures permutation(a[..],old(a[..]))
{
var j := i;
while j  rel> 0 && a[j-1]  rel> a[j] 
	invariant 0 rel<= j rel<= i && sortedBetween(a,0,j) && sortedBetween(a,j,i+1);
	invariant 1 rel< j+1 rel<= i ==> a[j-1] rel<= a[j+1];
	invariant permutation(a[..],old(a[..]));
	{
	a[j-1],a[j] := a[j],a[j-1];
	j := j-1;
	}
}
\end{lstlisting}
\vspace{-8mm}
\caption{Bubble Sort}
\label{bubblesort}
\end{figure} 
\vspace{-3mm}
\noindent{To} complete the program as given in Figure \ref{bubblesort}, we should enrich the specification of the method \D{bubbleSort} with the permutation property.
For that, we firstly should add a predicate expressing when a sequence of elements is a permutation of another sequence of elements. 
%Note that an array is an object which is different from the sequence of elements that this object contains. 
In Dafny, the sequence of elements  in an array \D{a} is denoted by \D{a[..]}, which is equivalent to 
\D{a[0..a.Length]}. Dafny sequences are an immutable value type.
Value types can be stored in fields (i.p. var) on the heap, and
used in real code in addition to specifications. Variables that
contain a value type can be updated to have a new value of
that type. Dafny also has multisets as inmutable value type. 
The built-in unary function \D{multiset} gives the multiset conversion of a sequence. 
\begin{lstlisting}[numbers=left, stepnumber=3, firstnumber=1]
predicate permutation (a:seq<int>,b:seq<int>)
{ multiset(a) == multiset(b) }
\end{lstlisting}
Now, we can add the ensures clause: 
\D{permutation(a[..],old(a[..]))} to the method 
\D{bubbleSort}. The expression \D{old(E)} stands for the value of the expression \D{E} when evaluated on entry to a method.
Consequently, it makes sense to use \D{old} in postconditions, and also in assertions (i.p. invariants) to refer to the entry value of a method parameter.
Then, we should add also the fact \D{permutation(a[..],old(a[..]))} to the invariant, but then it should be also a postcondition of the method \D{bubbleStep} and, for that, it should be preserved also by the loop in this method. 
We add all these facts and Dafny proves all them.

\section{Termination metrics}
\label{termination}
In this section we mainly give some hints on termination proofs, while introducing some other Dafny basic features, mainly datatypes and ghost variables. 
Dafny sets out to prove termination of all loops and of all recursion among methods and functions by means of \D{decreases} annotations.
A decreases annotation specifies an expression whose value is compared for successive loop iterations and for caller and callee.  Termination is ensured whenever the successive values become strictly smaller according to some well-founded order. 
Dafny has rules for guessing terminations metrics. For example, Dafny has successfully guessed a metric for every of the previous methods in this paper. In Section \ref{basic} we have already explained the guessed metric for function \D{factorial} and method \D{computeFactorial}.
The interested reader can check that Dafny also guessed 
\D{decreases k - i} for the while loop in
the method \D{compute5f},
\D{decreases a.Length - i} for the while loop in
the method \D{bubbleSort}, and
\D{decreases j - 0} (that is, \D{j}) for the while loop in
the method \D{bubbleStep}.
Though the most common metrics are of type integer, other types of expressions also work, including e.g. (finite) sequences (whose well-founded order Dafny defines to be proper-prefix ordering). In particular, tuples are very useful as termination metrics. Dafny compare tuples lexicographically.
%In the case of function \D{exp}, 
%Dafny guessed the pair of parameters \D{x,e} as metric, since it uses lexicographical order, the only condition is that \D{e} must be non-negative. Hence, the precondition is crucial for termination purposes:\footnote{Note that \D{e} is indeed the component 1 of the expression. In Dafny the first number is always 0.} 
%\includegraphics[scale=0.35]{termination-1.pdf}
If the guessed metric is not fine enough for proving termination, Dafny system asks the user to provide one through the hover text:\\ [2mm]
\includegraphics[scale=0.5]{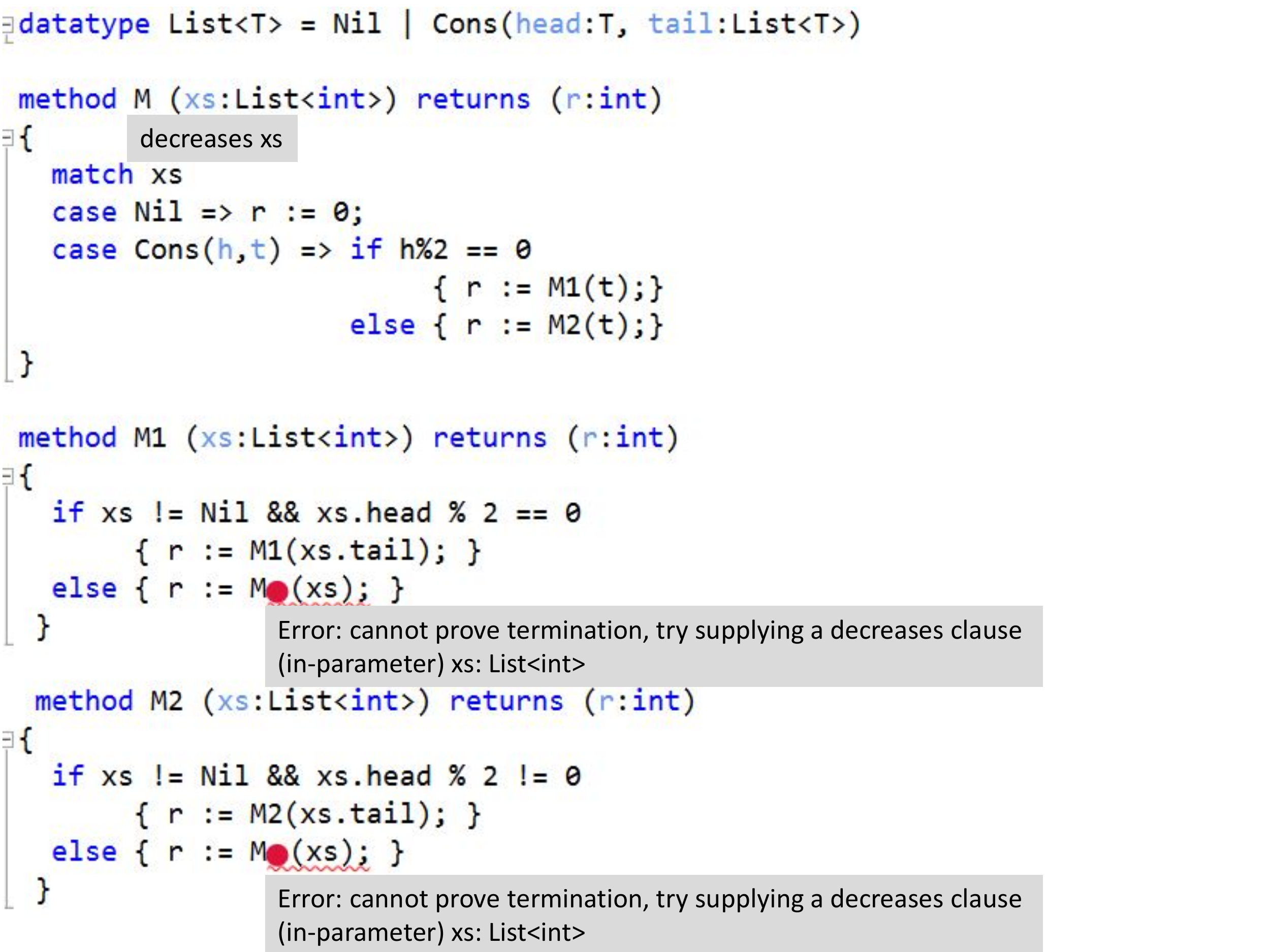}\\
The above example is really a skeleton or scheme\footnote{To carry out a real task, the code should be completed with more actions after the method calls, but we would like to concentrate on termination of this kind of mutual recursion.}
of mutual recursion that is interesting from the point of view of termination.
It deals with the usual algebraic/inductive \D{datatype}
of polymorphic lists\footnote{In the three methods type variable \D{T} is instantiated to \D{int}.} 
based on contructors  \D{Nil} and \D{Cons}
where, we have also declared the two usual destructors: \D{head} and \D{tail}$\!$. In methods \D{M1} and \D{M2} we use these destructors and the \D{if} statement to   implement recursion, whereas the method \D{M} uses the \D{match} statement. 
For the method \D{M}, Dafny guessed that the in-parameter \D{xs} could serve to prove termination, provided that the other two methods also terminate. But the same expression  \D{xs}, which is the only guessing in \D{M1} and \D{M2}, is not fine enough to prove termination.
In both cases, the \D{else} path is the issue, where the method \D{M} is called without modifying the parameter \D{xs}.
The most direct solution is to use a pair of integer expressions as metric. 
For the method \D{M} we write \D{decreases xs, 1}, whereas 
\D{decreases xs, 0} serves in the other two methods.
Therefore, the first expression is a variable and the second is a constant (either 0 or 1), so a strict decrease happens if the value of the variable strictly goes down, or if the variable remains unchanged and the caller has the 1 and the callee has the 0.

Dafny \emph{ghost} entities are used only during
verification; the compiler omits them from the executable
code, hence they can be used to help Dafny to complete a proof without jeopardizing execution cost. 
Ghost variables are useful when a non-computed value \D{v} has some interesting property that would allow
to prove the required property, but \D{v} is not really needed in
the real code. 
The following example is a bit contrived but illustrative nevertheless. The method (scheme) \D{CreateArray} generates an array \D{a}, using only one index \D{i}, to generate components in the following order: \D{a[0],a[n-1],a[1],a[n-2],a[2],a[n-3],a[3],...} where \D{n == a.Length}.\\

\hspace{-8mm}
\includegraphics[scale=0.35]{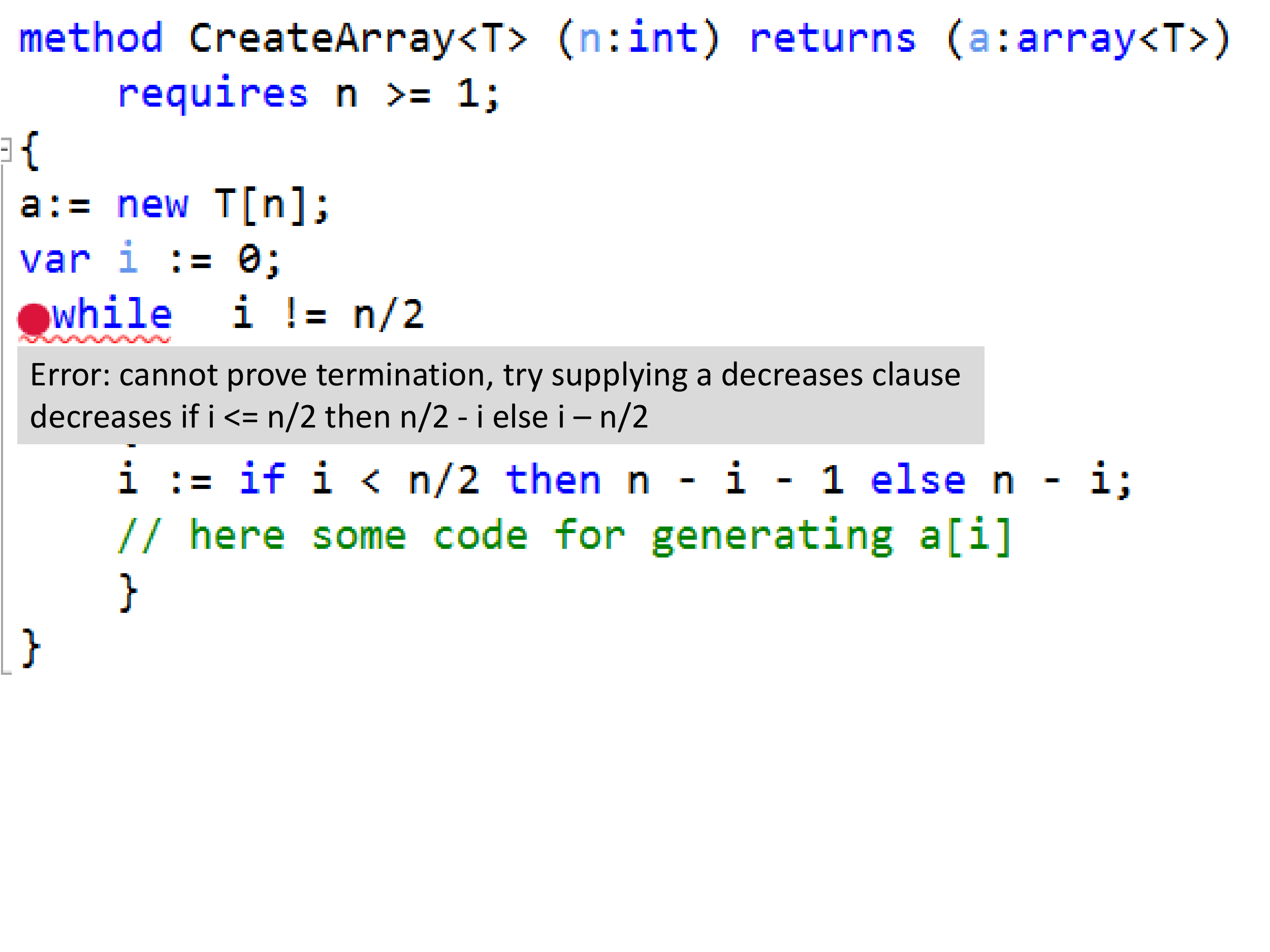} \\[-1.8cm]

It should be noted that \D{T} is a type variable and that the method  name is followed by \D{<T>} to denote that the method depends on that variable i.e. is polymorphic.

Dafny heuristics yields an expression (see the above hover text) that indeed decreases but not strictly. A suitable solution for proving termination is to introduce some kind of counter as a ghost variable. That is, we declare and initialize:
\D{ghost var c := 1;}. Then, the ghost variable \D{c}
should be incremented at each iteration. Hence, we can envisage that \D{n-c} is a good metric for termination. For that, we should add an invariant stating that \D{ 0 rel<= i rel<= n && c rel<= n}. However, \D{c rel<= n} cannot be proved to be maintained by the loop. For helping Dafny to prove \D{c rel<= n}, we should add some invariant property that relates the ghost variable \D{c} with \D{i} and \D{n}. There are many possible invariants that  enable Dafny to prove termination. 
One possible invariant is depicted in line 9 below:\\
\begin{lstlisting}[numbers=left, stepnumber=3, firstnumber=1]
method CreateArray<T> (n:int) returns (a:array<T>)
  requires n rel>= 1;
{
a:= new T[n];
var i := 0;
ghost var c := 1;
while i != n/2
	invariant 0 rel<= i rel<= n && c rel<= n;
	invariant c == 2*(n-i) || c == 2*i+1;
	decreases n-c;
	{ 
	i := if i rel< n/2 then n - i - 1 else n - i;
	// here some code for generating a[i]
	c := c + 1;
	}
}
\end{lstlisting}
Other interesting applications of ghost variables include to simplify specifications and  to specify class invariants in OO programming.

\section{Conclusion} 

Dafny is one of the many state-of-the-art tools that are currently being used in academic and industrial projects for the construction of reliable software.
On the one hand, the Dafny language is easy to learn though it incorporates most of the good features of modern programming and specification languages.
On the other hand, Dafny enables assertional proofs of correctness written as part of the program text.
Correctness proofs explain the reasons of the  program correctness. While proof ingredients are provided by the user,  the proof steps themselves are carried out automatically by the verifier.

I hope that this tutorial contributes to encourage the teaching of tool-supported formal verification issues in Software Engineering curricula.
I also hope that research
and developments will continue aiming at making formal methods more applicable. 

The development of formal verification tools (DFVT) is promoting many interesting research areas --such as invariant generation, quantification handling, etc.-- closely related to the field of automated reasoning (AR). Both areas, DFVT and AR, are mutually benefiting from their achievements.\\

\noindent{{\bf Acknowledgments.}
I am very grateful to the anonymous referees for their constructive comments.}

% Bibliography
\input{Lucio.bbl}

% \bibliographystyle{plain}
% \bibliography{NatMerge}
\end{document}

%% file: Lucio.bbl
\providecommand{\urlalt}[2]{\href{#1}{#2}}
\providecommand{\doi}[1]{doi:\urlalt{http://dx.doi.org/#1}{#1}}